# Astro2020 Science White Paper

# Synergies Between Galaxy Surveys and Reionization Measurements

**Thematic Areas:** ☐ Planetary Systems  ☐ Star and Planet Formation
☐ Formation and Evolution of Compact Objects  ☒ Cosmology and Fundamental Physics
☐ Stars and Stellar Evolution  ☐ Resolved Stellar Populations and their Environments
☒ **Galaxy Evolution**  ☐ Multi-Messenger Astronomy and Astrophysics

**Principal Author:**
Name: Steven Furlanetto
Institution: University of California, Los Angeles
Email: sfurlane@astro.ucla.edu
Phone: (310) 206-4127

**Co-authors:**
Adam Beardsley (Arizona State University), Chris L. Carilli (Cavendish Laboratory, NRAO), Jordan Mirocha (McGill University), James Aguirre (University of Pennsylvania), Yacine Ali-Haimoud (New York University), Marcelo Alvarez (University of California Berkeley), George Becker (University of California Riverside), Judd D. Bowman (Arizona State University), Patrick Breysse (CITA), Volker Bromm (University of Texas at Austin), Philip Bull (Queen Mary University of London), Jack Burns (University of Colorado Boulder), Isabella P. Carucci (University College London), Tzu-Ching Chang (JPL), Hsin Chiang (McGill University), Joanne Cohn (University of California Berkeley), Frederick Davies (University of California Santa Barbara), David DeBoer (University of California Berkeley), Mark Dickinson (NOAO), Joshua Dillon (University of California Berkeley), Olivier Doré (JPL, California Institute of Technology), Cora Dvorkin (Harvard University), Anastasia Fialkov (University of Sussex), Steven Finkelstein (University of Texas Austin), Nick Gnedin (Fermilab), Bryna Hazelton (University of Washington), Daniel Jacobs (Arizona State University), Kirit Karkare (University of Chicago/KICP), Leon Koopmans (Kapteyn Astronomical Institute), Ely Kovetz (Ben-Gurion University), Paul La Plante (University of Pennsylvania), Adam Lidz (University of Pennsylvania), Adrian Liu (McGill University), Yin-Zhe Ma (University of KwaZulu-Natal), Yi Mao (Tsinghua University), Kiyoshi Masui (MIT Kavli Institute for Astrophysics and Space Research), Matthew McQuinn (University of Washington), Andrei Mesinger (Scuola Normale Superiore), Julian Munoz (Harvard

University), Steven Murray (Arizona State University), Aaron Parsons (University of California Berkeley), Jonathan Pober (Brown University), Brant Robertson (University of California Santa Cruz/Institute for Advanced Study), Jonathan Sievers (McGill University), Eric Switzer (NASA Goddard Space Flight Center), Nithyanandan Thyagarajan (NRAO), Hy Trac (Carnegie Mellon University), Eli Visbal (Flatiron Institute), Matias Zaldarriaga (Institute for Advanced Study)

**Abstract:** The early phases of galaxy formation constitute one of the most exciting frontiers in astrophysics. It is during this era that the first luminous sources reionize the intergalactic medium – the moment when structure formation affects every baryon in the Universe. Here we argue that we will obtain a complete picture of this era by combining observations of galaxies with direct measurements of the reionization process: the former will provide a detailed understanding of bright sources, while the latter will constrain the (substantial) faint source population. We further describe how optimizing the comparison of these two measurements requires near-infrared galaxy surveys covering large volumes and retaining redshift information and also improvements in 21-cm analysis, moving those experiments into the imaging regime.

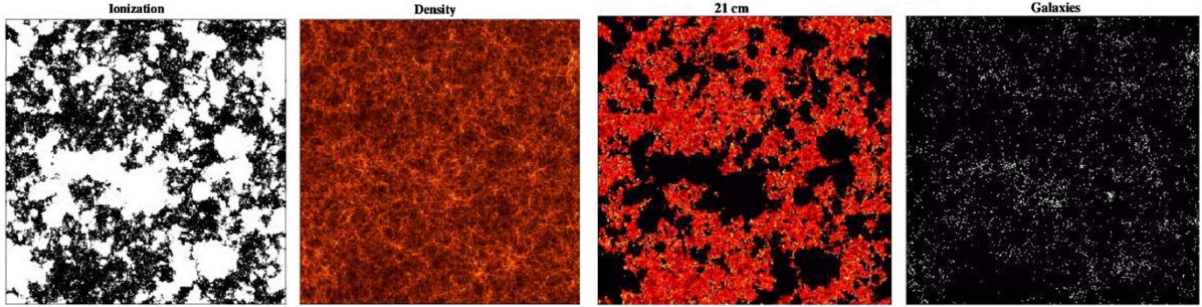

Fig. 1: **The relation between the galaxy and 21-cm fields contains an enormous amount of information about reionization and about the sources themselves.** Slices through a numerical simulation of reionization. The panels show the dark matter density field (second from left), the galaxy locations (right; white points), the ionization field (left; white regions are ionized) and the 21-cm brightness temperature (second from left; black regions have zero signal). Note how the galaxies and neutral gas are anticorrelated. From [1].

## I. INTRODUCTION

The formation and evolution of the first generations of galaxies, during the first billion years of the Universe's history, is a key goal of a number of forthcoming observational facilities. These galaxies host the transformation from the very first "Population III" stars to the more conventional processes governing star formation for the rest of the Universe's history, and it is likely at this time that the key scaling properties of galaxies (and their supermassive black holes) were established. Studying this era is difficult – most importantly, the galaxies are extraordinarily faint, challenging to find even with the advanced instruments available in the next decade. Galaxy surveys will undoubtedly chase these sources and provide powerful new measurements of their properties, but they will offer only an incomplete view of the early population.

This is also the era in which galaxy formation affects *every* baryon in the Universe through the large-scale radiation fields generated by these sources. X-rays (from stellar remnants or active galactic nuclei) heat the intergalactic medium, while ionizing photons eventually *reionize* the gas. So far, the reionization process can only be constrained indirectly (e.g., [2-8]), but low-frequency radio telescopes will observe the process directly in the next decade (e.g., [9-10]).

The details of this reionization process are intimately tied to the galaxy population, providing an exciting opportunity to improve our understanding of the first galaxies by combining measurements of reionization with galaxy surveys (see Fig. 1; [1, 11-17]). In this white paper, we will describe the science enabled by this combination and describe how both kinds of surveys can be optimized to maximize the science return of both efforts.

## II. OPPORTUNITIES FOR THE NEXT DECADE

In the near future, we expect substantial improvements in both galaxy and reionization measurements. JWST will provide an unprecedented window into high-redshift galaxy populations. It, together with ELTs, will begin to constrain the detailed properties of those galaxies through spectral measurements. However, the galaxies are so faint that even with an "ultradeep" survey, JWST will see only about half the light at $z\sim 10$ and only 10% at $z\sim 15$, at least according to simple models of the evolving galaxy population (e.g., [18-19]).

Large-area surveys, such as WFIRST, will complement JWST by providing censuses of the bright galaxy population across large cosmological scales. Variations across large (>50 Mpc) scales are expected to be much more important at high-redshifts than at later times, because a variety of radiative feedback processes (including X-ray heating and reionization) act on such scales and fundamentally alter the process of galaxy formation (e.g., [20-22]). Some of these processes may even leave relic signatures on later studies of the galaxy population [23]. However, sparsely-sampled, shallow measurements of the galaxy population will not be sensitive to such effects because feedback is strongest in small, young galaxies. Deep surveys are one viable approach, as is *intensity mapping* (low-resolution galaxy observations tracing the cumulative emission of the entire population; see white papers by Kovetz et al.; La Plante et al.).

Over the next several years, low-frequency radio telescopes will also measure the 21-cm signal from neutral hydrogen in the intergalactic medium during and before the reionization era. Instruments like the Hydrogen Epoch of Reionization Array (HERA) [10] will measure fluctuations in the 21-cm signal, primarily due to ionized bubbles in the IGM or variations in other radiation fields. These measurements will require a great deal of modeling to relate them to IGM and galaxy properties.

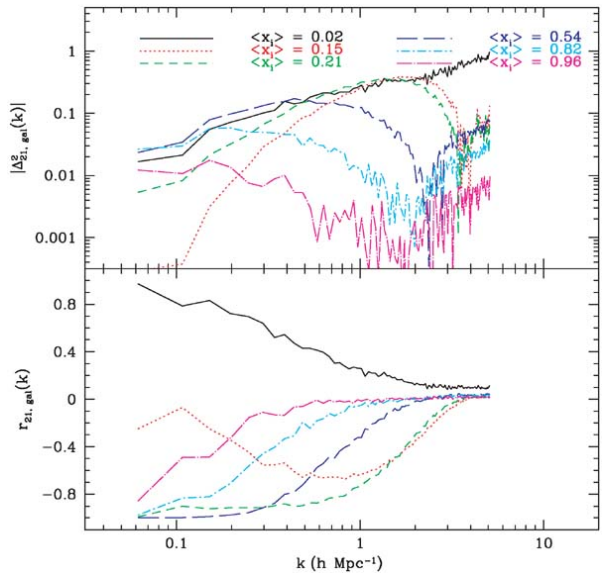

Fig. 2: **The statistical cross-correlation of the galaxy and 21-cm fields reveals ionized bubble sizes during reionization.** The upper panel shows the cross-correlation power spectrum at several different times in a numerical simulation of reionization (labeled by the ionization fraction $\langle x_i \rangle$). The bottom panel shows the cross-correlation coefficient for the same sequence of models. Note how, through most of reionization, the coefficient transitions from anti-correlation on large scales ($r$=-1) to no correlation ($r$=0). The scale at which this transition occurs is the typical size of an ionized bubble. From [1].

### III. SCIENCE ENABLED BY CROSS-CORRELATION

Fortunately, 21-cm measurements and galaxy surveys are highly complementary: the radiation fields that set the 21-cm signal are sensitive to the *integrated* emission of the entire galaxy population, providing leverage on the faint galaxies missed by direct surveys [20-22]. Meanwhile, the comparison with a known galaxy population breaks many of the degeneracies in inferences about the 21-cm signal [1]. Here, we describe several specific aspects of this comparison, and in Section IV we describe how future experiments can be optimized for these efforts.

One key galaxy parameter that has proven extremely difficult to measure directly is the escape fraction of ionizing photons. Measurements at lower redshifts have detected ionizing photons emerging from a fraction of galaxies, but no consensus has emerged on properties across the entire population (e.g., [24-29]). Precise measurements of the timing of reionization provide a census of the *total* ionizing emission from galaxies. A comparison with galaxy surveys therefore constrains the escape fraction to an unprecedented degree [30-31]; the measurement can be further improved by comparison to intensity mapping studies of faint galaxy populations.

Moreover, because galaxy and 21-cm surveys are complementary (measuring source and IGM properties, respectively), even statistical cross-correlations can break other important degeneracies. For example, one of the key reionization parameters is the typical size of the ionized bubbles as a function of time [32-33]. This information is contained in the 21-cm fluctuations, albeit only in a model-dependent way (e.g., [34]). However, a statistical cross-correlation between the 21-cm signal and galaxy locations reveals the scale *directly* (see Fig. 2; [1]). At a fixed ionized fraction, the most important factor determining this scale is the typical mass of the galaxies driving reionization, which can therefore be measured by a cross-correlation without requiring detailed galaxy modeling.

Although the first 21-cm measurements will be statistical, much more information about the details of galaxies and reionization can be extracted by direct comparisons of galaxy populations in different ionization environments. Beardsley et al. [16] have shown that, even in low signal-to-noise 21-cm observations, some neutral and ionized regions can be identified (see Fig. 3). That would enable the investigation of several key science questions about how the first phases of galaxy evolution depend on environment:

(1) Ionized regions grow around galaxy overdensities, which also means that they should host larger, older systems. Detailed galaxy spectra comparing populations in large and small ionized bubbles can probe these expectations and test how galaxies grew over time.
(2) Reionization likely raises the IGM temperature by at least an order of magnitude. This raises the Jeans mass and suppresses star formation in small galaxies, which may be important even for understanding dwarf galaxies at the present day (e.g., [35]) . Deep galaxy searches in ionized regions can test whether and how this suppression occurs [36-38].
(3) Although the *average* galaxy driving reionization can be characterized by a statistical cross-correlation, measuring the detailed trends requires isolating the contributions of different classes of galaxies. Studying the galaxy populations as a function of ionization environment allows us to do just that. In other words, it enables a direct test of the total contribution of galaxies to reionization as a function of their underlying properties (such as mass). Similarly, a comparison to Lyman-α emitting galaxies allows a study of the physics of that line.
(4) By directly measuring the source parameters in ionized regions, we can also constrain the role of IGM absorption in a less model-dependent way. In effect, once the source properties are known in detail, residual variations in the ionization environment must be driven by IGM absorption. This absorption is likely dominated by small-scale clumps, the equivalent of lower-redshift Lyman-limit systems, which cannot be directly studied at that time because of the extreme opacity of the Lyman-α forest [39-43].

(5) Ionized bubbles by definition contain galaxies. In the earliest phases of reionization, when (bright) galaxies are very rare, one of the principal challenges in studying them will be finding them: at $z$~15, even relatively bright sources will be out of reach of (relatively) shallow, large-area surveys, but they will be so rare that small field-of-view instruments (like JWST and ELTs) will have difficulty finding them [18,19]. Ionized regions detected with 21-cm telescopes can therefore act as galaxy-finders for very high-redshift surveys.

(6) Finally, the largest challenge in making 21-cm measurements is separating the weak cosmological signal from extremely bright foregrounds. Cross-correlation with a signal known to be at high redshifts (like the galaxy field) makes this procedure easier and will lend credence to the 21-cm measurement [11,44].

So far, we have considered the possibilities of comparing the galaxy population to the ionization state of the surrounding IGM. However, the 21-cm signal also depends on the temperature of the surrounding gas: that dependence is very weak if the IGM is heated well beyond the CMB temperature, but if the heating has not yet saturated, the thermal state of the IGM can also be measured [20-22]. In the most likely scenario, that temperature is determined by X-rays from stellar remnants and/or active galactic nuclei in these early galaxies. Probes of the heating era will be just as illuminating as the comparison to ionization features, allowing us to study the growth of accreting black holes in early Universe as a function of galaxy properties (see also white paper by Mirocha et al.).

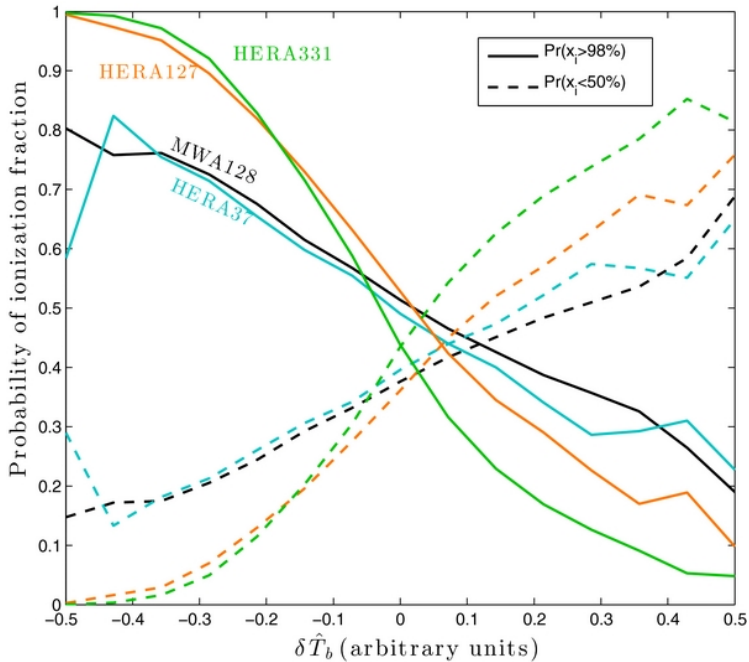

Fig. 3: **Existing 21-cm instruments can recover some ionized and neutral regions with high fidelity.** We show the probability that a (2.8 Mpc)$^3$ pixel in a 21-cm map is completely ionized (solid lines) or mostly neutral (dashed lines) for several different existing or planned instruments. Extrema in the maps can be reliably partitioned into these categories. Further work developing imaging technology for these instruments will greatly expand the fraction of pixels that can be identified as ionized or mostly neutral. From [16].

### IV. KEY SCIENCE ADVANCES

Unsurprisingly, in practice the comparison between 21-cm measurements and galaxy surveys is challenging [1,11]. The 21-cm signal is extremely weak. The current generation of instruments hopes to measure its properties statistically, by making low signal-to-noise measurements but using large volumes (over tens of square degrees) to average out the noise. 21-cm telescopes like HERA are further limited by foregrounds (which are several orders of magnitude brighter than the cosmological signals): to prevent contamination, these telescopes largely ignore angular information, focusing on the better-measured radial structure [45-46].

Moreover, the radiation backgrounds to which the 21-cm signal is sensitive induce features on large scales: the typical ionized bubbles are ~50 Mpc during much of reionization, spanning ~25 arcminutes (e.g., [32,47]). Surveying a representative volume during reionization therefore requires extremely large surveys (covering at least several square degrees) but must also map galaxies in the radial direction. This is much larger than the fields of view of most NIR instruments. The optimal cross-correlation is therefore with a **large-area, deep galaxy survey that has precise redshift information.**

The more detailed tests outlined in Section III require the identification of ionized and neutral regions at high confidence. While even low signal-to-noise observations can find some such regions [16], improving the analysis framework of 21-cm surveys will make the task much easier. This requires efficiently removing contamination from foregrounds so that at least some angular modes can be recovered – **fully leveraging the 21-cm galaxy relation requires making images with 21-cm telescopes**. Such improvements will also dramatically improve statistical cross-correlations, because they allow access to many more modes.

**Analysis of these synergistic observations requires sophisticated models that capture the relevant physical effects**. This is not an easy task: early generations of galaxies likely have all the physical complexity of their descendants (and possibly more, to account for the transition from primordial to "normal" star formation), but their evolution is much more closely coupled to their large-scale environment thanks to feedback from large-scale radiation fields like the ionizing background. Simulations are beginning to capture some of these effects, but given the massive uncertainties in galaxy formation parameters at these times, full simulations are likely too costly for statistical comparisons with data. Continued development of galaxy formation and reionization theory, on all levels from analytic models to simulations, will be necessary to optimize the extraction of astrophysical constraints from these observations.

## V. CONCLUSIONS

The reionization era is a compelling frontier for studying both the first generations of galaxies and the intergalactic medium. But a complete understanding of it will only be possible with complementary observations of both the source population and their effects on the IGM. An effective combination of galaxy surveys and the 21-cm background requires substantial preparation, including the development of wide-field survey capabilities in the near-infrared, image analysis improvements for 21-cm telescopes, and targeted techniques to identify the most interesting regions of the IGM for follow-up. The potential science return of these combinations is enormous.